\documentclass[runningheads]{llncs}

\usepackage[T1]{fontenc}

\emergencystretch=\maxdimen
\usepackage{graphicx}
\begin{document}
\title{UAST: Unicode Aware Saṃskṛta Transliteration}
\author{Dhruvil Dave\orcidID{0000-0003-0616-7277} \and Aneri Dalwadi\orcidID{0000-0002-1771-1576}}

\institute{}

\maketitle

\begin{abstract}
\textit{Devanāgarī} is the writing system that is adapted to various languages like \textit{Saṃskṛta}. International Alphabet of Sanskrit Transliteration (IAST) is a transliteration scheme for the romanization of Sanskrit language. IAST makes use of diacritics to represent various characters. On a computer, these are represented using the Unicode standard, which differs from how the Sanskrit language behaves at a very fundamental level. This results in an issue that is encountered while designing typesetting software for Devanāgarī and IAST. We hereby discuss the problems and provide a solution that solves the issue of incompatibilities between various transliteration and encoding schemes. The base implementation that should be used is available at \texttt{https://github.com/dhruvildave/uast.rs}. Another implementation that extends UAST to around $10$ scripts is available at \texttt{https://github.com/aneri0x4f/uast-cli} and \texttt{https://github.com/dhruvildave/uast} .

\end{abstract}

\newpage
\section{Handling Unicode Characters in IAST}
\subsection{Pros and Cons of IAST}\label{iast-pc}

\paragraph{}
International Alphabet of Sanskrit Transliteration (IAST) is a popular transliteration scheme used by the Saṃskṛta language for the romanization of the Devanāgarī script that has seen adaptation among scholars.

\paragraph{}
IAST has many benefits:
\begin{itemize}
    \item Closely follows the original scripts.
    \item Unambiguous.
    \item Intuitive.
    \item Case insensitive.
\end{itemize}

\paragraph{}
IAST uses various diacritics with letters. When it comes to typesetting Saṃskṛta with IAST, we have to
use non-ASCII Unicode characters which creates a problem of having to use non-standard key bindings
and/or packages to input those characters. Figuring out those key bindings and setting up IAST input
has been a pain point for scholars and students alike. For example, depending on the operating system, package, or editor that is being used, ā can be entered through \texttt{Ctrl + a}, \texttt{Alt + a}, or any other key binding in that case. This greatly hinders portability.

\subsection{UAST-IO syntax for handling Unicode}\label{uast-io}

\paragraph{}
We suggest a new syntax, \textbf{UAST-IO}, for representing the characters that use diacritics that are used in IAST:

\begin{itemize}
    \item Any non-ASCII character should be entered between \texttt{/x/} as per Table \ref{tab-unicode} where \texttt{x} should be replaced by the ASCII replacement for the corresponding Unicode character. We use SOLIDUS (\texttt{U+002F}), also known as the forward slash.
    \item All other ASCII characters should be entered as-is.
\end{itemize}

\begin{table}
\centering
\caption{Unicode diacritic to ASCII mapping}\label{tab-unicode}
\begin{tabular}{|l|c|c|}
\hline
\textbf{Unicode Name} & \textbf{Code Point} & \textbf{ASCII replacement}\\
\hline
LATIN SMALL LETTER A WITH MACRON & \texttt{0101} & \texttt{a}\\
LATIN SMALL LETTER A WITH TILDE & \texttt{00E3} & \texttt{au}\\
LATIN SMALL LETTER I WITH MACRON & \texttt{012B} & \texttt{i}\\
LATIN SMALL LETTER U WITH MACRON & \texttt{016B} & \texttt{u}\\
LATIN SMALL LETTER R WITH DOT BELOW & \texttt{1E5B} & \texttt{r}\\
LATIN SMALL LETTER R WITH DOT BELOW AND MACRON & \texttt{1E5D} & \texttt{ru}\\
LATIN SMALL LETTER L WITH DOT BELOW & \texttt{1E37} & \texttt{l}\\
LATIN SMALL LETTER L WITH DOT BELOW AND MACRON & \texttt{1E39} & \texttt{lu}\\
LATIN SMALL LETTER L WITH LINE BELOW & \texttt{1E3B} & \texttt{ll}\\
LATIN SMALL LETTER T WITH DOT BELOW & \texttt{1E6D} & \texttt{t}\\
LATIN SMALL LETTER D WITH DOT BELOW & \texttt{1E0D} & \texttt{d}\\
LATIN SMALL LETTER N WITH TILDE & \texttt{00F1} & \texttt{n}\\
LATIN SMALL LETTER N WITH DOT ABOVE & \texttt{1E45} & \texttt{nu}\\
LATIN SMALL LETTER N WITH DOT BELOW & \texttt{1E47} & \texttt{nl}\\
LATIN SMALL LETTER S WITH ACUTE & \texttt{015B} & \texttt{su}\\
LATIN SMALL LETTER S WITH DOT BELOW & \texttt{1E63} & \texttt{sl}\\
LATIN SMALL LETTER M WITH DOT BELOW & \texttt{1E43} & \texttt{m}\\
LATIN SMALL LETTER H WITH DOT BELOW & \texttt{1E25} & \texttt{h}\\
\hline
\end{tabular}
\end{table}

Table \ref{tab-unicode} values have been derived as follows:

\begin{itemize}
    \item If a diacritic appears above of a character, the character is appended with a \texttt{u} where \texttt{u} stands for "upper". If a diacritic appears below of a character, the character is appended with a \texttt{l} where \texttt{l} stands for "lower".

    \begin{itemize}
        \item LATIN SMALL LETTER S WITH ACUTE becomes \texttt{su}.
        \item LATIN SMALL LETTER N WITH DOT ABOVE becomes \texttt{nu}.
        \item LATIN SMALL LETTER S WITH DOT BELOW becomes \texttt{sl}.
        \item LATIN SMALL LETTER N WITH DOT BELOW becomes \texttt{nl}.
    \end{itemize}

    \item For characters that only have a single diacritic, we mention the character directly as it is unambiguous by default.

    \begin{itemize}
        \item LATIN SMALL LETTER I WITH MACRON becomes \texttt{i}.
        \item LATIN SMALL LETTER U WITH MACRON becomes \texttt{u}.
        \item LATIN SMALL LETTER D WITH DOT BELOW becomes \texttt{d}.
        \item LATIN SMALL LETTER T WITH DOT BELOW \texttt{t}.
        \item LATIN SMALL LETTER M WITH DOT BELOW \texttt{m}.
        \item LATIN SMALL LETTER H WITH DOT BELOW becomes \texttt{h}.
    \end{itemize}

    \item For diacritics with a clash, frequency of usage guides selection of replacement. The higher the frequency, the lower the length of replacement.

    \begin{itemize}
        \item LATIN SMALL LETTER A WITH MACRON becomes \texttt{a} and LATIN SMALL LETTER A WITH TILDE becomes \texttt{au}.
        \item LATIN SMALL LETTER R WITH DOT BELOW becomes \texttt{r} and LATIN SMALL LETTER R WITH DOT BELOW AND MACRON becomes \texttt{ru}.
        \item LATIN SMALL LETTER L WITH DOT BELOW becomes \texttt{l}, LATIN SMALL LETTER L WITH DOT BELOW AND MACRON becomes \texttt{lu}, and LATIN SMALL LETTER L WITH LINE BELOW becomes \texttt{ll}.
    \end{itemize}

    \item Remaining diacritics are mapped to simplest possible representations that are closest to its IAST counterparts.

    \begin{itemize}
        \item LATIN SMALL LETTER N WITH TILDE becomes \texttt{n}.
    \end{itemize}
\end{itemize}

\paragraph{}
This mapping is intended for replacement of non-ASCII characters only. Thus, \texttt{ḍha} should be written
as \texttt{/d/ha} and not \texttt{/dh/a} as only the former would be a correct mapping while latter would be an invalid case as there is no mapping defined for \texttt{dh} but only \texttt{d}.

\paragraph{}
The benefits that arise from doing this is that now we have a standard way of representing unicode glyphs without having to worry about keyboard layouts and key bindings.

\paragraph{}
There are two main reasons why we use a new syntax, as inspired by \LaTeX, and not resort to other techniques that are adopted by other schemes and encodings.

\begin{itemize}
    \item Using a mix of lowercase and uppercase would make the scripting and typesetting case-sensitive which tends to be error-prone, harm readability, and involve lot of extra invisible key-pressing in form of \texttt{Shift} or \texttt{Caps Lock}. Our system makes typing those extra characters completely explicit and remove the problem of case-sensitivity altogether.
    \item The only new character which gets introduced is SOLIDUS (\texttt{U+002F}) which would explicitly demark the character boundary. Adding other non-alphanumeric characters like AMPERSAND (\texttt{U+0027}) or FULL STOP (\texttt{U+002E}). A suitable competitor to forward slash would be LOW LINE (\texttt{U+005F}) but it would involve pressing \texttt{Shift} key. We could equivalently use HYPHEN-MINUS (\texttt{U+002D}) but we reserve it for a special purpose that we present in the next section.
\end{itemize}

\paragraph{}
Here are a few examples to exemplify the above principles:

\begin{table}
\centering
\caption{\textit{Bhagavadgītā} 15.15 (UAST-IO)}\label{tab-andy}
\begin{tabular}{|c|c|}
\hline
\textbf{IAST} & \textbf{UAST-IO}\\
\hline
\texttt{sarvasya cāhaṃ hṛdi saṃniviṣṭo} & \texttt{sarvasya c/a/ha/m/ h/r/di sa/m/nivi/sl//t/o}\\
\texttt{mattaḥ smṛtirjñānamapohanaṃ ca} & \texttt{matta/h/ sm/r/tirj/n//a/namapohana/m/ ca}\\
\texttt{vedaiśca sarvairahameva vedyo} & \texttt{vedai/su/ca sarvairahameva vedyo}\\
\texttt{vedāntakṛdvedavideva cāham} & \texttt{ved/a/ntak/r/dvedavideva c/a/ham}\\
\hline
\end{tabular}
\end{table}

\begin{table}
\centering
\caption{A comparison of various transliteration schemes or encodings for Saṃskṛta.}\label{tab-comparison0}
\begin{tabular}{|r|l|l|l|}
\hline
\textbf{IAST} & \texttt{rāmaḥ} & \texttt{kṛṣṇaḥ} & \texttt{prajñā}\\
\hline
\textbf{Harvard-Kyoto} & rAmaH & kRSNaH & prajJA\\
\textbf{SLP1} & rAmaH & kfzRaH & prajYA\\
\textbf{Velthuis} & raama.h & k.r.s.na.h & praj$\sim$naa\\
\textbf{ITRANS} & rAmaH & kRRiShNaH & praj$\sim$nA\\
\textbf{UAST-IO} & \textbf{\texttt{r/a/ma/h/}} & \textbf{\texttt{k/r//sl//nl/a/h/}} & \textbf{\texttt{praj/n//a/}}\\
\hline
\end{tabular}
\end{table}

\newpage
\section{Devanāgarī and Unicode}
\subsection{Problems with existing systems}

\paragraph{}
The previous section dealt with the problem of handling the input of Unicode characters when using IAST. The situation does not improve when we move from IAST to Devanāgarī. We have the same problems as above.

\paragraph{}
We already have many systems that make use of ASCII characters to input the Devanāgarī script. A very popular system is the Harvard-Kyoto convention. It uses a mix of lowercase and uppercase ASCII characters. In this case, \texttt{rāma} in IAST would become \texttt{rAma} in Harvard-Kyoto.

\paragraph{}
Comparison of various systems is best done by formalizing the requirements for a good typesetting system.

\begin{itemize}
    \item Case insensitive.
    \item Uses only ASCII characters.
    \item Does not involve invisible key-pressing like \texttt{Shift} or \texttt{Caps Lock}.
    \item Follow IAST.
    \item The character entered should be intuitive and follow the pronunciation of the language.
\end{itemize}

\paragraph{}
All other encodings fall short in at least one of the criteria. But the problems amplify.

\subsection{Unicode Aware Saṃskṛta Transliteration}

\paragraph{}
We can use the UAST-IO syntax (Section \ref{uast-io}) and extend it further. We suggest the following system for typesetting Devanāgarī:

\begin{itemize}
    \item The typesetting should follow IAST.
    \item Any unicode diacritics to be entered in IAST should follow the system presented in Section \ref{uast-io} and all ASCII characters should simply be entered as is.
    \item Put an explicit HYPHEN-MINUS (\texttt{U+002D}) to represent a \textit{virāma} (\texttt{U+094D}). Thus, unlike IAST, \textit{k} would become \textit{k-} to signify a lack of inherent vowel.
    \item \textit{a} after a consonant is redundant but supported. So, \textit{kamala} and \textit{kml} would both mean the same.
    \item To place an explicit vowel instead of a vowel sign, follow it with a REVERSE SOLIDUS (\texttt{U+005C}). It is also used to explicitly mark the character boundary. \textit{dh} would be parsed as \textit{dha}. To treat them individually, place \textit{a} like \textit{dah} which would be parsed as \textit{daha}.
    \item The typesetting should be case-insensitive. Thus, both \textit{p} and \textit{P} should be treated in the same way.
\end{itemize}

\paragraph{}
We call this the Unicode Aware Saṃskṛta Transliteration (UAST) system. Any system that implements UAST will be able to unambiguously transpile it to IAST and Devanāgarī using an algorithm with linear space and time complexity. The primary purpose of UAST is to bridge the gap between how the typesetting of IAST and Devanāgarī is done.

\paragraph{}
The following benefits arise from using UAST:

\begin{itemize}
    \item It is case-insensitive.
    \item All the key-presses are explicit. There are no uses of keys like \texttt{Shift} or \texttt{Caps~Lock}.
    \item The only non-alphanumeric characters are SOLIDUS, REVERSE SOLIDUS, and FULL STOP.
    \item It is highly intuitive, as it does not deviate from IAST and follows character sounds.
\end{itemize}

\subsection{Design Decisions}

Several decisions were involved and inspired the design of UAST.

\subsubsection{Following IAST}

The main problem with certain encodings and keyboard-layouts is that they map many Devanāgarī characters to key-bindings that are non intuitive. They distance themselves from IAST. IAST is widely used for many reasons:

\begin{itemize}
    \item Its mapping is unambiguous and exact.
    \item Related sounds have related mappings. This makes IAST a very logical and intuitive mapping to use. As a result, IAST bridges the gap between Saṃskṛta and English phonetics.
    \item Follows the original script as-is.
\end{itemize}

\paragraph{}
This makes it arguably the most popular transliteration scheme used by scholars and students alike. Since IAST closely follows the language, it would make sense that the Devanāgarī typesetting also follows IAST to prevent confusion and remove ambiguity. We believe that a person trying to learn or use a language and interact with it via a typesetting system should be more concerned with "what to write" rather than "how to write".

\paragraph{}
The more an encoding or transliteration system distances itself from IAST, the less intuitive it becomes. Other systems work well in isolation, but only add to more confusion when it comes to dealing with them alongside IAST.

\subsubsection{Avoiding mix of lowercase and uppercase characters}

There are major disadvantages involved with the use of both lowercase and uppercase characters. The first is slowing of the reading and writing speeds (\cite{babayigit}). Consider the following example. \texttt{pitā} in IAST would be written as \texttt{pitA} in Harvard-Kyoto and \texttt{pit/a/} in UAST. The first three characters remain the same in both. The difference occurs at the end. Ignoring any user-specific or system-specific changes, there are two possible sequences to enter \textit{A} after \textit{t}:

\begin{itemize}
    \item \texttt{Shift + a}
    \item \texttt{Caps Lock, a, Caps Lock}
\end{itemize}

The former involves two key presses, while the latter does three. Thus, to write \texttt{pitA}, we would have to do five or six key presses. For writing \texttt{pit/a/}, there is no uppercase involved. Thus, it would be completed in six key presses. So, at the cost of additional key presses, we end up with two major benefits:

\begin{itemize}
    \item All the key-presses are now explicit. Infact, since we are now using all lowercase characters on a standard ASCII keyboard, number of \texttt{key-presses = length of string}. So, a user writes what a user sees. The errors arising from this are no longer a problem.
    \item The problem of case-sensitivity is completely eliminated. In conventions like Harvard-Kyoto, changing a character to lowercase even by mistake would incur a significant penalty on accuracy and precision of transliteration.
\end{itemize}

\subsubsection{Avoiding use of non-ASCII Unicode diacritics}

The main problem involved with IAST is to enter non-ASCII Unicode diacritics. There are many problems, as discussed in Section \ref{iast-pc} involved in entering non-ASCII Unicode diacritics as used in IAST. It might involve finding a specific package or even manually setting up key bindings. This raises the problem of portability. There is no standard syntax for Devanāgarī or even IAST in case of absence of Unicode diacritics on the keyboard. In UAST, we address this problem using the solutions suggested in Section \ref{iast-pc}.

\subsection{Explicit virāma and Choice of character set}

\paragraph{}
Saṃskṛta language and computers have a difference that is very fundamental to the whole problem of encoding, transliteration, and typesetting.

\begin{figure}
\centering
\includegraphics[width=200pt, height=200pt]{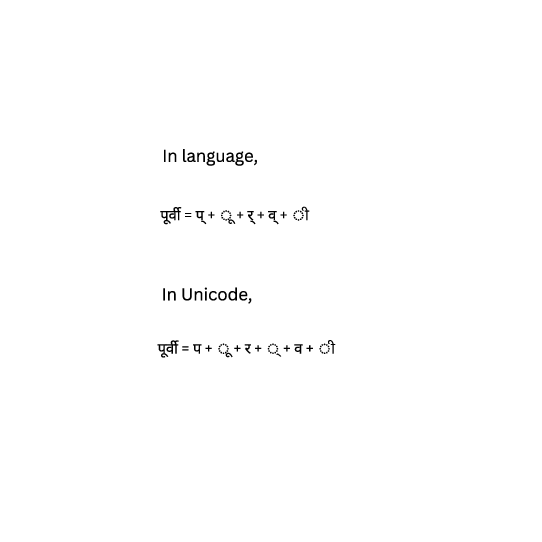}
\caption{Comparing \textit{pūrvī} for language vs computer.} \label{fig1}
\end{figure}

\paragraph{}
Saṃskṛta language is written using IAST or Devanāgarī script. Computers use Unicode encoding. In Saṃskṛta, the consonants are inherently without a vowel and vowels including \textit{a} are added later on. But this is not the case for Unicode. There is a very subtle but very important distinction. In Unicode Standard, consonants are inherently with an \textit{a} and \textit{virāma} is added later to signify the absence of any vowel. Refer to Fig. \ref{fig1}

\paragraph{}
This means that Unicode \cite{unicode2024} is fundamentally incompatible with the way language and IAST function. There is no explicit representation of the virāma in IAST as it is not needed after all. But then parsing IAST and converting it to Devanāgarī means that we have to look for the additional and missing \textit{a}. For example, on pressing \textit{k}, IAST assumes that the IAST equivalent of \textit{k} is entered but Unicode assumes that the IAST equivalent of \textit{ka} is entered. Now, a person using IAST will press \textit{a}. At this point, for the user, the IAST equivalent of ka should have been entered. But what we end up with is the IAST equivalent of \textit{kaa}. This occurs frequently with every consonant that IAST has to fight with Unicode for removal of that extra a to make things work as per standards. This addition and removal of characters makes it completely unintuitive to use IAST directly while typesetting Devanāgarī. This is a problem because at this point we are directly working with unicode glyphs of Devanāgarī rather than a romanised representation. It will not be a problem if IAST or even Devanāgarī functioned in isolation. Things get more confusing when going to both of IAST and Devanāgarī as various encodings add new characters that aren't a part of IAST which leads to unintuitive key-bindings in Devanāgarī.

\paragraph{}
To resolve this issue, UAST prefers the use of an explicit virāma to bridge this very gap between Devanāgarī/IAST and Unicode. Since IAST is already the standard to work with, the entire character set of UAST is the same as that of IAST with added \textit{-} for a virāma as it very much resembles a dash. So under UAST, when \textit{k} is entered, the IAST equivalent of \textit{ka} will be entered to take Unicode into account more closely and be in the center of the spectrum between the two. This makes \textit{a} redundant. This little detail can be very easily parsed, and it now comes with the added benefit that since the inherent vowel is now explicit, we can easily use UAST to directly parse it into Devanāgarī directly and to IAST with very little to no modification. UAST uses no diacritics and hence there is no problem of missing characters, nonstandard key bindings, or case sensitivity.

\paragraph{}
This makes UAST the single typesetting system and solving multiple encoding problems to represent Sanskrit in computers by mixing the good parts of both IAST and Unicode while being
intuitive, fast, and easy to learn, use, and remember at the same time.

%
%
%
\newpage
\bibliographystyle{splncs04}

\end{document}